\documentclass[aps,onecolumn,nopacs,nofootinbib,floatfix,superscriptaddress]{revtex4}
\usepackage{graphicx}
\usepackage{amsfonts}
\usepackage{amssymb}
\usepackage{amsbsy}
\usepackage{hyperref}
\usepackage{amsmath}
\usepackage{mathrsfs}
\usepackage{latexsym}
\usepackage{natbib}
\usepackage{bm}
\usepackage{subfigure} 
\usepackage{color}
\usepackage{wasysym}
\usepackage{mathbbol}
\usepackage{bigints}
\allowdisplaybreaks
\usepackage[normalem]{ulem}
\usepackage[dvipsnames]{xcolor}
\usepackage{multirow}
\usepackage{physics}
\usepackage{comment}
\usepackage{calligra}

\definecolor{napiergreen}{rgb}{0.16, 0.5, 0.0}

\begin{document}

\title{Mach's principle in atomic transitions}

\author{Subhajit Barman}
\email{subhajitbarman.phy@gmail.com}
\affiliation{Physics and Applied Mathematics Unit, Indian Statistical Institute, 203 B.T. Road, Kolkata 700108, India}
\affiliation{Centre for Strings, Gravitation and Cosmology, Department of Physics, Indian Institute of Technology Madras, Chennai 600036, India}

\author{Bibhas Ranjan Majhi}
\email{bibhas.majhi@iitg.ac.in}
\affiliation{Department of Physics, Indian Institute of Technology Guwahati, Guwahati 781039, Assam, India}

\begin{abstract}

We investigate the atomic transition probabilities in atom-mirror set-ups that are in circular motion. In one scenario, the atom is in circular motion inside a static cylindrical mirror. 
In the other scenario, the cylindrical mirror rotates around its central axis while the atom remains static. We report structural similarity in the atomic transition probabilities between these two cases -- these probabilities are equivalent upon interchanging the field frequencies between the two scenarios. We interpret such an observation as a semi-classical phenomenon analogous to the classical Mach's principle.

\end{abstract}

\date{\today}

\maketitle
 
\section{Introduction}\label{sec:Introduction}

The similarity between semiclassical gravity phenomena like the Hawking effect \cite{hawking1975} and the Unruh effect \cite{Unruh:1976db} has long been attributed to the classical similarity between the two scenarios due to Einstein's equivalence principle (EEP) \cite{Paunkovic:2022flx} -- the trajectories of a particle under gravity are indistinguishable locally from those for a free particle viewed with respect to an accelerated frame. In recent times, there has been renewed interest in further investigating and comparing the quantum responses of these observers, which are in classically equivalent scenarios \cite{Lammerzahl:1996se, Viola:1996de, Zych:2015fka, Pipa:2018bui, Paunkovic:2022flx} and those in analogous relative classical motions \cite{Svidzinsky:2018jkp, Fulling:2018lez, Sonego:2003zh}. For instance, in \cite{Svidzinsky:2018jkp}, the equivalence between the transition probabilities of two two-level atoms is demonstrated, one of which is in uniform acceleration relative to a static mirror, and the other atom is static while the mirror accelerates uniformly. Notably, this investigation is carried out in $(1+1)$ dimensions, and in $(3+1)$ dimensions \cite{Kumawat:2024kul}, there are subtle features in the atomic transitions that differentiate the two scenarios. At the same time, there are also investigations to understand the same between atom-mirror pairs placed between the Minkowski spacetime and its future or past light cone region \cite{Barman:2024dql}. Taking insight from the understanding of EEP, a similarity in the coordinate transformations between these scenarios (the Minkowski spacetime and its light-cone regions) and the interior of a Schwarzschild black hole is observed, and subsequently, the equivalence of the quantum phenomenon is also studied in \cite{Barman:2024dql}. The investigation is further extended to understand the semi-classical nature of the field mode and the detector, as well as the role of the spacetime dimensionality in this equivalence \cite{Kumawat:2026oro}. \vspace{0.1cm}

The pertinent question is whether any of these predicted equivalences in the quantum phenomenon can be verified in an experimental set-up in the near future. In this regard, several studies \cite{Rosi:2017ieh, Geiger:2018xwr, Tino:2020nla, Onofrio:2025ixt} have explored different possible directions. It is to be noted that the detection of an atomic transition, described in the previous paragraph, requires very high acceleration \cite{Stargen:2021vtg, Arya:2024qke, Ghosh:2024mqy, Barman:2024jpc}, which is beyond the scope of the present technology. This work aims to explore an experimentally viable possibility, atom-mirror set-ups in circular motion, to test a similar equivalence for a quantum system. Our choice of set-ups in circular motion is motivated by the fact that they are more convenient from the practical point of view as compared to linear accelerated motion \cite{Bell:1982qr, Davies:1996ks, Korsbakken:2004bv, Gutti:2010nv, Louko:2017emx, Biermann:2020bjh, Barman:2022utm, Peng:2024pfy}, and there have been significant developments in the experimental front in this direction \cite{Cohen:2021axm, Zhang:2022ree}. The experimental viability of this set-up also facilitates investigating the analogue of classical Mach's principle in the transition probabilities of quantum detectors.\vspace{0.1cm}

Classically, the environment seen from a frame in circular motion is equivalent to the rotating environment with respect to a static frame. This idea was challenged and discussed initially in Newton's bucket experiment \cite{newton1687principia} and then in the basic tenet of Mach's principle \cite{barbour1995mach, Das:2012uh}. In Newton's bucket experiment, in one scenario, a bucket half-filled with water is in circular motion, and the observer is stationed at the centre, and in the other scenario, the bucket is kept static at the centre, and the observer is in circular motion with the same radius and angular momentum as the first scenario. The shape of the water inside the bucket is expected to reveal whether the bucket or the observer is in circular motion, thereby distinguishing between two classical scenarios that appear equivalent in terms of the relative motion between the bucket and the observer. At the same time, Mach's principle says the inertia of an object can only be determined with respect to a truly inertial object, such as the large-scale distribution of matter in the rest of the universe. Therefore, the understanding of the shape of water inside a bucket as a determinant of inertia may itself be flawed, as all of our observations are limited to local frames. 
{Here, we aim to develop an understanding of the overall situation by investigating its analogue in the semi-classical regime.} 
In a similar spirit, recently one study has been done to compare the active and passive rotations \cite{Facundes:2025xcc}. 
{Through this semiclassical analogue of Mach’s principle, we aim to investigate how environmental motion influences quantum field states and particle creation, and whether equivalent effects arise in static environments for quantum probes following specific trajectories. We further examine the equivalence of these phenomena in non-relativistic and relativistic regimes, potentially offering insights into gravitational effects on quantum systems through Einstein’s Equivalence Principle (EEP), which relates local gravitational fields to uniformly accelerated frames.}
\vspace{0.1cm}

We consider a pair of atom-mirror\footnote{The mirror acts as a perfect reflecting boundary for the field modes.} set-ups where the atoms represent the quantum systems. These set-ups are in two different configurations, which are as follows.
\begin{enumerate}
    \item[I.] In the first scenario, the atom is in circular motion inside a static cylindrical boundary. We investigate the atomic transition probability in this scenario.
    \item[II.] In the second scenario, the atom is kept static, but the cylindrical boundary is in circular motion around its central axis, and we study the atomic transition probability.
\end{enumerate}
One can notice that these two scenarios are analogous in terms of the relative motions between the atoms and the boundary. We obtain the atomic transition probabilities in these two scenarios and check whether there is any apparent similarity between the two at the quantum level to reflect on the classical similarity. In order to investigate that we also express these probabilities in terms of the static and rotating field mode frequencies as measured by a rotating and static observer, respectively. Our observations indicate that the atomic transition probabilities for a single field mode corresponding to the two scenarios have similar structures and are equivalent when the field mode frequencies of the two cases are interchanged. Similar observations are also reported in our investigation of the integrated version of transition probability rates over all field mode frequencies. Here, we would like to mention that in \cite{DePaola:2000cv} where the authors shed light on the quantum version Mach’s principle with atomic detectors interacting with the field modes in the Trocheries-Takeno vacuum state, corresponding to a rotating frame, does not refer to equivalent detector transition rates between different scenarios. At the same time, while the equivalence in the field vacua is studied between static and rotating frames in the context of analogue gravity \cite{Jannes:2015fwa}, it is observed to hold true for low energies and fails only in high energy regimes. These previous investigations make our present work all the more interesting.\vspace{0.1cm}

The present manuscript is organized in the following manner. In Sec. \ref{sec:Setup-Coords} we elucidate the set-up for investigating the atomic transition probabilities. In this section, we also obtain the scalar field modes in static and rotating frames of Minkowski spacetime. In Sec. \ref{sec:atomic-Rot-exctn} we estimate the transition probability when the atom is in circular motion inside a static cylindrical boundary. Subsequently, in Sec \ref{sec:atomic-St-exctn} we evaluate the transition probability of a static atom inside a rotating cylindrical boundary. In Secs. \ref{sec:atomic-Rot-exctn} and \ref{sec:atomic-St-exctn} we estimate both the single mode and mode integrated atomic transition probabilities. Finally, in Sec. \ref{sec:discussion} we draw attention to key observations and conclude with a discussion of their implications.\vspace{0.1cm}

We shall work with units such that $c=1$, where $c$ signifies the speed of light.

\section{The set-up and the field mode solution}\label{sec:Setup-Coords}
In this section, we elucidate the formulation and set-up to investigate the atomic transition probability in the presence of a cylindrical boundary. In particular, we specify the scenarios where either the atom is in motion, and the boundary is static, or the atom is static, and the boundary is in motion. We will also obtain the field mode solutions in the appropriate frames of these different scenarios, i.e., in the static and rotating frames.

\subsection{The set-up}
We consider a Unruh-DeWitt detector, modelled as a two-level atom, and interacting weakly with the background real scalar quantum field. Excitation of the field mode will give rise to a particle, which will be called a scalar photon. We consider the initial atom-field state to be $|\Psi \rangle$, and the Hamiltonian to be $\hat{H} = \hat{H}_{A} + \hat{H}_{F} + \hat{H}_{I}$. It is to be noted that $\hat{H}_{A}$, $\hat{H}_{F}$, and $\hat{H}_{I}$ respectively denote the free-atomic, free-field, and interaction Hamiltonians. The interaction Hamiltonian, see \cite{Svidzinsky:2018jkp}, is modeled as 
\begin{eqnarray}\label{eq:Int-Hamiltonian}
    \hat{H}_{I}(\tau) &=& \hbar\,g\,\Big(\phi_{\nu}[t(\tau),z(\tau)]\,\hat{a}_{\nu}+h.c.\Big)\,\big(e^{-i\,\omega\,\tau}\hat{\sigma}+h.c.\big)~,
\end{eqnarray}
where $\hbar$ denotes the reduced Planck's constant, $g$ the coupling strength between the background field and the atom, $\phi_{\nu}$ the field mode, $\hat{a}_{\nu}$ the lowering operator corresponding to the field, $\hat{\sigma}$ the atomic lowering operator, $\omega$ is the atomic energy gap, and $\tau$ is the proper time in the atom's frame. We further consider that the atom is initially in its ground state $|\omega_{0}\rangle$, while $|\omega_{1}\rangle$ is the atom's excited state. At the same time, we consider the field to be in the ground state $|0\rangle$. We would like to obtain the atomic transition probability when we also have simultaneous emission of a single photon. One can obtain this transition probability \cite{Svidzinsky:2018jkp} in the lowest order of coupling strength with the help of the above interaction Hamiltonian as
\begin{eqnarray}\label{eq:Atom-ExProb}
    \mathcal{P}^{ex}_{\nu}(\omega) &=& \frac{1}{\hbar^2} \bigg|\int_{-\infty}^{\infty}\,d\tau \, \langle 1_{\nu},\omega_{1}|\hat{H}_{I}(\tau)|0,\omega_{0}\rangle\bigg|^2~\nonumber\\
    ~&=& g^2\,\bigg|\int_{-\infty}^{\infty}\,d\tau \, \phi^{\star}_{\nu}[t(\tau),z(\tau)]\,e^{i\,\omega\,\tau}\bigg|^2~.
\end{eqnarray}
The above expression is obtained considering $g\ll 1$, and keeping only the leading order term in the Dyson series. It should be mentioned that the above transition probability corresponds to a specific field mode of frequency $\nu$. In our subsequent analysis, we will utilize this expression to obtain the atomic transition probabilities in different scenarios. \vspace{0.1cm}

Here we would like to point out that an atomic transition probability, when a sum over all field modes is considered, can also be obtained from expression \eqref{eq:Atom-ExProb}. In this regard, we consider the interaction Hamiltonian to be $\hat{H}_{I} = \hbar\,g\,\Phi[x(\tau)]\,\big(e^{-i\,\omega \,\tau} \hat{\sigma} +h.c.\big)$, where $\Phi(x)=\int d^{3}k\, [\phi_{k}\,\hat{a}_{k} + \phi^{\star}_{k}\, \hat{a}^{\dagger}_{k}]$ represents the entire field decomposed in terms of individual field modes $\phi_{k}$. Then the field mode integrated atomic transition probability, see \cite{Barman:2024dql}, will be given by
\begin{eqnarray}\label{eq:Atom-ExProb-MdIndp}
    \mathcal{P}^{ex}(\omega) &=& \int d^{3}k ~ \mathcal{P}^{ex}_{\nu}(\omega) = g^2\,\int d^{3}k\, \bigg|\int_{-\infty}^{\infty}\,d\tau \, \phi^{\star}_{\nu}[t(\tau),z(\tau)]\,e^{i\,\omega\,\tau}\bigg|^2~.
\end{eqnarray}
{It is straightforward to see that the coupling strengths $(g)$ appearing in Eqs. \eqref{eq:Atom-ExProb} and \eqref{eq:Atom-ExProb-MdIndp} must have different dimensions for the corresponding transition probabilities to be dimensionless. This difference arises from the distinct forms of the initial interaction Hamiltonians considered in the two cases.}
It should be noted that in the above expression $\nu$ depends on the different components of the field wave vector $\vec{k}$. This transition probability may diverge for specific atomic motions as it is estimated over infinite time, and thus it is convenient to define transition probability rates. In particular, in our present setting, one can define the transition probability rate, see \cite{Barman:2022utm}, as
\begin{eqnarray}\label{eq:Rex-MdIndp}
    \mathcal{R}^{ex}(\omega) &=& \frac{\mathcal{P}^{ex}(\omega)}{\lim_{\tilde{T}\to \infty}\, \int_{-\tilde{T}}^{\tilde{T}} d\eta}~.
\end{eqnarray}
In our subsequent analysis, we shall investigate both the single-mode transition probability of \eqref{eq:Atom-ExProb} and the mode integrated transition probability rate \eqref{eq:Rex-MdIndp} in different scenarios, and compare them to talk about the quantum equivalence between different scenarios.

\subsection{The field mode solution}

\begin{figure*}
\includegraphics[width=7.8cm]{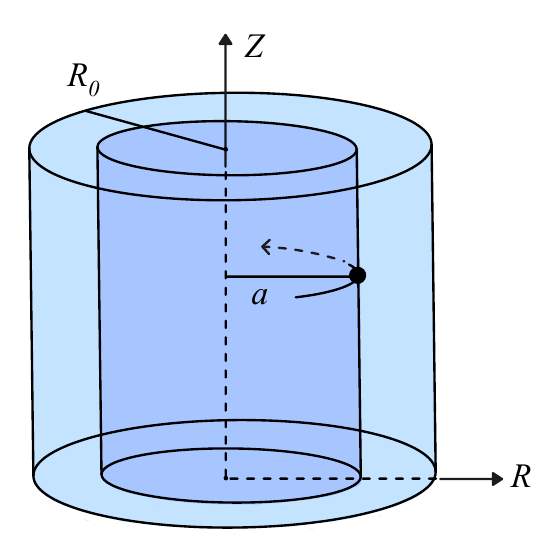}
\hskip 30pt
\includegraphics[width=7.8cm]{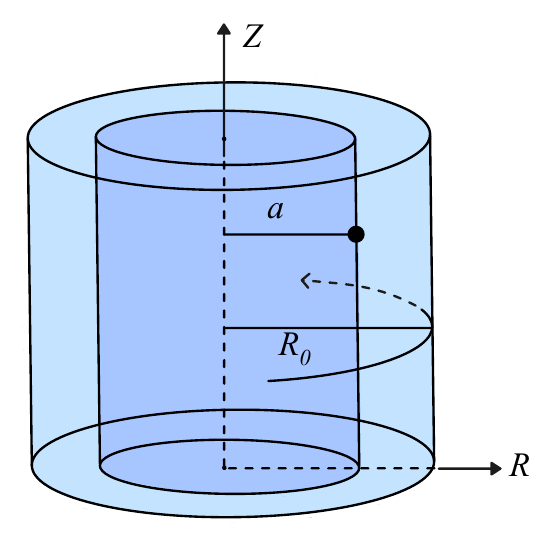}
\caption{In the above figure, we have provided the schematic diagrams for the two different scenarios. On the left, the atom is in circular motion maintaining a radial distance $a$ from the central axis, i.e., the $Z-$axis. At the same time, the mirror is a static cylinder of radius $R_{0}$. On the right, the atom is static at a radial distance $a$ from the central axis. In the second scenario, the mirror is cylindrical, with radius $R_{0}$, and rotates about its central axis.}
    \label{fig:2-Scenarios}
\end{figure*}

In this part of the section, we discuss the coordinates that facilitate constructing set-ups with an atom and a cylindrical mirror in different circular motions corresponding to different scenarios. In particular, in one scenario, the atom is in circular motion inside a static cylindrical mirror, and in the other scenario, the atom is static while the mirror, along with the environment, is in circular motion. See Fig. \ref{fig:2-Scenarios} to check the schematic diagrams depicting these different scenarios. We also obtain the field modes with these coordinates.

We consider the cylindrical polar coordinate system. The cylindrical polar coordinates associated with a static frame in Minkowski spacetime are $(T,R,\Phi,Z)$. At the same time, the coordinates associated with a frame in circular motion with angular velocity $\Omega$ are $(t,\rho,\varphi,z)$. The coordinate transformation \cite{Picanco:2020api, Barman:2022utm} from one frame to another is
\begin{eqnarray}\label{eq:coord-trns}
    T &=& \gamma\,t~,\nonumber\\
    ~R &=& \rho~,\nonumber\\
    ~\Phi &=& \varphi + \gamma\,\Omega\,t~,\nonumber\\
    ~Z &=& z~;
\end{eqnarray}
where $\gamma=1/\sqrt{1-v^2}$ is the Lorentz factor with $v$ denoting the linear speed of the frame in circular motion, i.e., $v=\Omega\,\rho=\Omega\,R$. The line element in these coordinates is given by
\begin{subequations}\label{eq:coords-st-rt}
\begin{eqnarray}\label{eq:coords-st}
    ds^2 &=& -dT^2+dR^2+R^2\,d\Phi^2+dZ^2~,\\
    ~&=& -dt^2+d\rho^2+\rho^2\,d\varphi^2+dz^2+2\,\gamma\,\Omega\,\rho^2\,d\varphi\,dt~.\label{eq:coords-rt}
\end{eqnarray}
\end{subequations}
We shall evaluate the field mode solutions with respect to the static and rotating frames. It is to be noted that the field modes for the inertial frame can be obtained by making $\Omega=0$ in the field modes of the rotating frame.

First, in the static frame, the Klein-Gordon (KG) equation of motion, $(1/\sqrt{-g})\,\partial_{a}(\sqrt{-g}\,g^{ab}\,\partial_{b}\Psi) = \mu^2\,\Psi$, corresponding to a massive scalar field $\Psi$ with mass $\mu$ is given by
\begin{eqnarray}\label{eq:KG-eom-St}
    \partial_{T}^{2}\Psi -\frac{1}{R} \partial_{R}(R\,\partial_{R}\Psi)-\frac{1}{R^2}\, \partial_{\Phi}^2\Psi-\partial_{Z}^2\Psi+\mu^2\,\Psi &=& 0~. 
\end{eqnarray}
As the metric components in the line-element of Eq. \eqref{eq:coords-st} are independent of coordinates $(T,\Phi,Z)$, we consider the field mode decomposition $U_{\mathcal{E}\,m\,k} (T,R,\Phi,Z) \sim \exp{-i\,\mathcal{E}\,T + i\,m\Phi + i\,k\,Z} \times \mathbb{R}(R)$. With this field mode decomposition, the KG equation of motion becomes
\begin{eqnarray}\label{eq:KG-eom-St-2}
    \frac{1}{R}\,\partial_{R} \big(R\,\partial_{R}\mathbb{R}\big)+\bigg[\chi^2-\frac{m^2}{R^2}\bigg]\mathbb{R} &=& 0~,
\end{eqnarray}
where $\chi^2=\mathcal{E}^2-k^2-\mu^2$. The solution to the differential equation \eqref{eq:KG-eom-St-2} can be obtained in terms of $\mathcal{J}_{m}(\chi\,R)$ and $Y_{m}(\chi\,R)$, where $\mathcal{J}_{m}(x)$ and $Y_{m}(\chi\,R)$ respectively signify the Bessel functions of the first and the second kind. However, only $\mathcal{J}_{m}(\chi\,R)$ is non-divergent at $R=0$, and thus to get a well-behaved solution of the field mode inside the cylindrical boundary we must have $\mathbb{R}(R)\sim \mathcal{J}_{m}(\chi\,R)$, see \cite{Letaw:1980ik}. Therefore, before normalization the field mode looks like $U_{\mathcal{E}\,m\,k} (T,R,\Phi,Z) \sim \exp{-i\,\mathcal{E}\,T + i\,m\Phi + i\,k\,Z} \times \mathcal{J}_{m}(\chi\,R)$. The inner product \cite{Picanco:2020api} between field modes $\psi_{i}$ and $\psi_{j}$ is defined as
\begin{eqnarray}\label{eq:Inr-prdct}
    \langle \psi_{i},\,\psi_{j} \rangle &=& i\int_{\Sigma} \sqrt{q}~d\Sigma_{a}\, \Big[\psi_{i}^{\star}\, n^{a}\psi_{j}-\psi_{j}\, n^{a}\psi_{i}^{\star} \Big]~,
\end{eqnarray}
where $q$ is the determinant of the metric induced on the hypersurface $d\Sigma_{a}$ and $n^{a}$ is the time-like Killing vector with respect to which the inner product is defined. From the Dirichlet's boundary condition of $U_{\mathcal{E}\,m\,k} (T,R_{0},\Phi,Z)=0$, we obtain the field mode solution as $U_{\mathcal{E}\,m\,k} (T,R,\Phi,Z) \sim \exp{-i\,\mathcal{E}\,T+i\,m\,\Phi+i\,k\,Z}\,\mathcal{J}_{m}(\chi_{mn}\,R)$, where $\chi_{mn}$ is the $n^{th}$ root of the $m^{th}$ Bessel function, i.e., it is the solution of $\mathcal{J}_{m}(\chi_{mn}\,R_{0}) = 0$ \cite{Picanco:2020api}. Then the normalized field mode solution on the fixed-$T$ hypersurface will be of the form (see \cite{Picanco:2020api, Stargen:2021vtg, Arya:2024qke})
\begin{eqnarray}\label{eq:FM-StMirror}
    U_{\mathcal{E}\,m\,k} (T,R,\Phi,Z) &=& \frac{\exp{-i\,\mathcal{E}\,T+i\,m\,\Phi+i\,k\,Z}\,\mathcal{J}_{m}(\chi_{mn}\,R)}{2\pi R_{0}\sqrt{\mathcal{E}}\,|\mathcal{J}'_{m}(\chi_{mn}\,R_{0})|}~.
\end{eqnarray}

Second, in the rotating frame, the Klein-Gordon (KG) equation of motion, $(1/\sqrt{-g})\,\partial_{a}(\sqrt{-g}\,g^{ab}\,\partial_{b}\Psi) = \mu^2\,\Psi$, corresponding to the massive scalar field $\Psi$ becomes
\begin{eqnarray}\label{eq:KG-eom-rt}
    \partial_{t}^{2}\Psi -\frac{\gamma^2}{\rho} \partial_{\rho}(\rho\,\partial_{\rho}\Psi)-\frac{1}{\rho^2}\, \partial_{\varphi}^2\Psi-\gamma^2\, \partial_{z}^2\Psi-2\,\Omega\,\gamma\, \partial_{t}\partial_{\varphi}\Psi+\gamma^2\,\mu^2\,\Psi &=& 0~. 
\end{eqnarray}
As the metric components in Eq. \eqref{eq:coords-rt} are independent of coordinates $(t,\varphi,z)$, we consider the field mode decomposition $u_{\varepsilon\,m\,k} (t,\rho,\varphi,z) \sim \exp{-i\,\varepsilon\,t+i\,m\varphi+i\,k\,z}\,\mathcal{R}(\rho)$. With this field mode decomposition, the KG-EOM becomes
\begin{eqnarray}\label{eq:KG-eom-rt-2}
    \frac{1}{\rho}\,\partial_{\rho} \big(\rho\,\partial_{\rho}\mathcal{R}\big)+\bigg[\kappa^2-\frac{m^2}{\rho^2}\bigg]\mathcal{R} &=& 0~,
\end{eqnarray}
where $\kappa^2=(m\Omega+\varepsilon/\gamma)^2-k^2-\mu^2$. The well-behaved solution to the above equation of motion of \eqref{eq:KG-eom-rt-2} is $\mathcal{R}(\rho)\sim \mathcal{J}_{m}(\kappa\,\rho)$. Thus before normalization the field mode looks like $u_{\varepsilon\,m\,k} (t,\rho,\varphi,z) \sim \exp{-i\,\varepsilon\,t + i\,m\varphi + i\,k\,z}\,\mathcal{J}_{m}(\kappa\,\rho)$. Here also if we consider the Dirichlet's boundary condition of $u_{\varepsilon\,m\,k} (t,R_{0},\varphi,z)=0$, we obtain the field mode solution as $u_{\varepsilon\,m\,k} (t,\rho,\varphi,z) \sim \exp{-i\,\varepsilon\,t+i\,m\,\varphi+i\,k\,z}\,\mathcal{J}_{m}(\kappa_{mn}\,\rho)$, where $\kappa_{mn}$ is the $n^{th}$ root of the $m^{th}$ Bessel function, i.e., it is the solution of $\mathcal{J}_{m}(\kappa_{mn}\,R_{0}) = 0$ and we have $\chi_{mn}=\kappa_{mn}$. Then, with the help of Eq. \eqref{eq:Inr-prdct}, the normalized field mode solution on the fixed-$t$ hypersurface can be obtained as
\begin{eqnarray}\label{eq:FM-StAtom}
    u_{\varepsilon\,m\,k} (t,\rho,\varphi,z) &=& \frac{\exp{-i\,\varepsilon\,t+i\,m\,\varphi+i\,k\,z}\,\mathcal{J}_{m}(\kappa_{mn}\,\rho)}{2\pi R_{0}\sqrt{\varepsilon}\,|\mathcal{J}'_{m}(\kappa_{mn}\,R_{0})|}~.
\end{eqnarray}
In our subsequent analysis, we shall use the above field mode solutions from Eqs. \eqref{eq:FM-StMirror} and \eqref{eq:FM-StAtom} to obtain the atomic transition probabilities.\vspace{0.1cm}

The energy carried by the static frame mode $U_{\mathcal{E}\,m\,k} (T,R,\Phi,Z)$ measured by a static observer is $\mathcal{E}$ as is evident from $ i\,\partial_{T}\, U_{\mathcal{E}\,m\,k} (T,R,\Phi,Z) = \mathcal{E}\, U_{\mathcal{E}\,m\,k} (T,R,\Phi,Z)$. At the same time, the same carried by the rotating frame mode $u_{\varepsilon\,m\,k} (t,\rho,\varphi,z)$ with respect to a static observer is $\varepsilon'=m\,\Omega+\varepsilon/\gamma$, which can be obtained by operating the energy operator $i\,\partial_{T}$ corresponding to the static observer on the field mode $u_{\varepsilon\,m\,k} (t,\rho,\varphi,z)$ as follows
\begin{eqnarray}\label{eq:Mode-energy-rel}
  i\,\frac{\partial}{\partial T}~u_{\varepsilon\,m\,k} (t,\rho,\varphi,z) = i\,\Big[\frac{\partial t}{\partial T}\frac{\partial}{\partial t}+\frac{\partial \varphi}{\partial T}\frac{\partial}{\partial \varphi}\Big]~u_{\varepsilon\,m\,k} (t,\rho,\varphi,z) = \Big[\frac{\varepsilon}{\gamma}+m\,\Omega\Big]~u_{\varepsilon\,m\,k} (t,\rho,\varphi,z)~.
\end{eqnarray}
{To derive the above expression, we have used the coordinate transformation given in Eq. \eqref{eq:coord-trns}.} 
Similarly, the energy carried by the static frame mode, as measured by an observer in circular motion, is $\mathcal{E}' = \gamma(\mathcal{E} - m\Omega)$, which is obtained from $i\,\partial_t\, U_{\mathcal{E}\,m\,k} (T,R,\Phi,Z) = \mathcal{E}'\, U_{\mathcal{E}\,m\,k} (T,R,\Phi,Z)$.

\section{Atom is in circular motion and mirror is static}\label{sec:atomic-Rot-exctn}

In this section, we consider the particular scenario when the atom is in circular motion inside a static cylindrical mirror. For a pictorial representation of this scenario, one can check the left schematic diagram in Fig. \ref{fig:2-Scenarios}. In particular, we consider that the atom is at radius $a$ inside the cylindrical mirror of radius $R_{0}$. In this scenario, the atomic trajectory is described by $T=\gamma\,\tau$, $R=a$, $\Phi=\gamma\Omega\tau$, and $Z=0$, with $\tau$ being the proper time of the atom. Then, with the help of the field mode solution \eqref{eq:FM-StMirror} corresponding to a static frame, we obtain the atomic transition probability of \eqref{eq:Atom-ExProb} as
\begin{eqnarray}\label{eq:Pex-StMirror}
    \mathcal{P}^{ex}_{\mathcal{E}}(\omega) &=& g^2\bigg|\int_{-\infty}^{\infty}d\tau~ U^{\star}_{\mathcal{E}\,m\,k} (T,R,\Phi,Z)\,e^{i\,\omega\,\tau}\bigg|^2~\nonumber\\
    ~&=& \frac{g^2}{4\pi^2 R_{0}^2\,\mathcal{E}}\,\bigg|\frac{\mathcal{J}_{m}(\chi_{mn}\,a)}{\mathcal{J}'_{m}(\chi_{mn}\,R_{0})}\bigg|^2\times \bigg|\int_{-\infty}^{\infty}d\tau~\,e^{i\,(\omega+\mathcal{E}\gamma-m\gamma\Omega)\,\tau}\bigg|^2~\nonumber\\
    ~&=& \frac{g^2}{R_{0}^2\,\mathcal{E}}\,\bigg|\frac{\mathcal{J}_{m}(\chi_{mn}\,a)}{\mathcal{J}'_{m}(\chi_{mn}\,R_{0})}\bigg|^2\times \big|\delta[\omega-\gamma(m\Omega-\mathcal{E})]\big|^2~.
\end{eqnarray}
{The Dirac delta distributions in the above expression arise from the eternal atom–field interaction assumed in Eq. \eqref{eq:Int-Hamiltonian}, reflecting infinite interaction time rather than physical divergences. This is also evident for a static atom (see Appendix \ref{Appn:Static-detector}). Although a finite-duration interaction can regularize these distributions via a switching function (see Appendix \ref{Appn:TP-mode-Regu}), it introduces dependence on the switching profile and timescale \cite{LSriramkumar_1996}, potentially obscuring effects due to atomic motion and the boundary. Since the mode-integrated transition rates are free of these distributions, the eternal-interaction framework provides a natural idealization for studying these effects.}

It is to be noted that the above expression is obtained in terms of field mode frequency $\mathcal{E}$, i.e., with respect to the static frame mode frequency as seen by a static observer. Here we would like to mention that this expression can be further simplified. For instance, one can identify $\gamma (\mathcal{E}-m\Omega)$ as $\mathcal{E}'$ the static frame mode frequency as seen by an observer in circular motion. Then the above expression for atomic transition probability becomes
\begin{eqnarray}\label{eq:Pex-StMirror-2}
    \mathcal{P}^{ex}_{\mathcal{E},\,\mathcal{E}'}(\omega) &=& \frac{g^2}{R_{0}^2\,\mathcal{E}}\,\bigg|\frac{\mathcal{J}_{m}(\chi_{mn}\,a)}{\mathcal{J}'_{m}(\chi_{mn}\,R_{0})}\bigg|^2\times \big|\delta[\omega+\mathcal{E}']\big|^2~.
\end{eqnarray}
This expression \eqref{eq:Pex-StMirror-2} will be particularly helpful in comparing the atomic transitions between different scenarios in our subsequent analysis.

\subsection{Field mode integrated transition rate}

\begin{figure*}
\includegraphics[width=7.9cm]{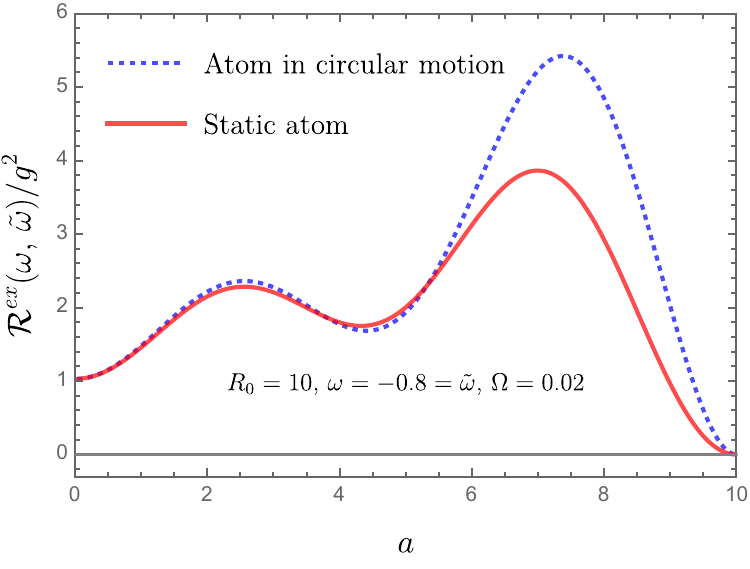}
\hskip 30pt
\includegraphics[width=7.9cm]{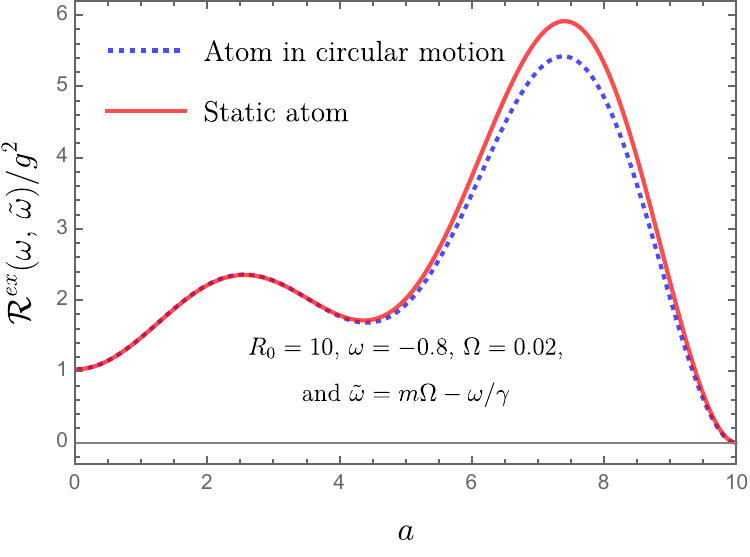}
\caption{In the above figure, we have plotted $\mathcal{R}^{ex}(\omega)/g^2$ and $\mathcal{R}^{ex}(\tilde{\omega})/g^2$, that respectively represent the atomic transition probability rates corresponding to two different scenarios, with respect to the radial position of the atom $a$. In particular, on the left, we consider the atomic energy gaps to be $\omega=\tilde{\omega}$. At the same time, on the right, we choose $\tilde{\omega}=m\,\Omega-\omega/\gamma$. We choose the maximum value in the numerical sum over $n$ to be $20$ and, over $m$, from $-20$ to $20$, beyond which we do not observe any deviation from the obtained results. We choose other parameters suitably. From both plots, one can notice that the transition rates are qualitatively equivalent. Moreover, in the right plot, when $\tilde{\omega}=m\,\Omega-\omega/\gamma$, the difference between $\mathcal{R}^{ex}(\omega)/g^2$ and $\mathcal{R}^{ex}(\tilde{\omega})/g^2$ do not change sign, and they are quantitatively the same as well for small values of $a$ (when $a\le 4$, as compared to the left plot).}
    \label{fig:RexVsa-w}
\end{figure*}

Next, we consider estimating the mode integrated atomic transition probability rate as prescribed in Eq. \eqref{eq:Rex-MdIndp}. With the help of Eqs. \eqref{eq:Atom-ExProb-MdIndp} and \eqref{eq:Rex-MdIndp}, and the previous expression \eqref{eq:Pex-StMirror}, we obtain this atomic transition rate as
\begin{eqnarray}\label{eq:Rex-StMirror1}
    \mathcal{R}^{ex}(\omega) &=& \frac{g^2}{R_{0}^2\,\big\{\lim_{\tilde{T}\to \infty} \, \int_{-\tilde{T}}^{\tilde{T}} d\eta\big\}}\,\sum_{m=-\infty}^{\infty}\sum_{n=1}^{\infty}\int_{-\infty}^{\infty} \frac{dk}{\mathcal{E}}\, \bigg|\frac{\mathcal{J}_{m}(\chi_{mn}\,a)}{\mathcal{J}'_{m}(\chi_{mn}\,R_{0})}\bigg|^2\, \big| \delta[\omega - \gamma(m\Omega-\mathcal{E})]\big|^2~.
\end{eqnarray}
For a massless scalar field $\mu=0$, we have the relation $\mathcal{E}=\sqrt{k^2+\chi^{2}_{mn}}$. We consider $f(k)=\mathcal{E}- (m\Omega-\omega/\gamma)$. Then one can simplify the above Dirac delta distribution as
\begin{eqnarray}
    \delta[\omega - \gamma(m\Omega-\mathcal{E})] &=& \frac{1}{\gamma}\, \sum_{i} \frac{\delta(k-k_{i})}{|f'(k_{i})|}~,
\end{eqnarray}
where $k_{i}=\pm \sqrt{(m\Omega-\omega/\gamma)^2-\chi^{2}_{mn}}$ and $|f'(k_{i})| = |k_{i}|/\sqrt{k^{2}_{i}+\chi^{2}_{mn}} = \sqrt{(m\Omega-\omega/\gamma)^2-\chi^2_{mn}}/(m\Omega-\omega/\gamma)$. Then the previous expression of the transition rate becomes
\begin{eqnarray}\label{eq:Rex-StMirror2}
    \mathcal{R}^{ex}(\omega) &=& \frac{4\,g^2}{R_{0}^2\,\gamma^2}\,\sum_{m=-\infty}^{\infty}\sum_{n=1}^{\infty}\, \bigg|\frac{\mathcal{J}_{m}(\chi_{mn}\,a)}{\mathcal{J}'_{m}(\chi_{mn}\,R_{0})}\bigg|^2\, \frac{(k^2_{i}+\chi^2_{mn})~\theta(m\Omega-\omega/\gamma-\chi_{mn})}{\mathcal{E}_{i}\,|k_{i}|^2}\nonumber\\
    ~&=& \frac{4\,g^2}{R_{0}^2\,\gamma^2}\,\sum_{m=-\infty}^{\infty}\sum_{n=1}^{\infty}\, \bigg|\frac{\mathcal{J}_{m}(\chi_{mn}\,a)}{\mathcal{J}'_{m}(\chi_{mn}\,R_{0})}\bigg|^2\, \frac{(m\Omega-\omega/\gamma)~\theta(m\Omega-\omega/\gamma-\chi_{mn})}{(m\Omega-\omega/\gamma)^2-\chi^2_{mn}}~.
\end{eqnarray}
This expression cannot be further simplified analytically, to our understanding. One can take the help of numerical methods to obtain the exact values of the above transition rate $\mathcal{R}^{ex}(\omega)$ for different system parameters. For instance, in Fig. \ref{fig:RexVsa-w}, we have plotted the above transition rate as a function of the radial position of the atom $a$ and compared it with the second scenario.

\section{Atom is static and mirror is in circular motion}\label{sec:atomic-St-exctn}

In this section, we consider the atom to be static at radius $a$ inside a cylindrical mirror of radius $R_{0}$. The mirror, along with the environment inside it, is in circular motion around its central $Z-$axis. One can check this set-up on the right diagram of Fig. \ref{fig:2-Scenarios}. The atomic trajectory in the coordinates of rotating frame is $t=\tau/\gamma$, $R=a$, $\varphi=-\Omega\,\tau$, and $Z=0$, with $\tau=T$ being the proper time of the atom. In this scenario, we consider the field mode solution \eqref{eq:FM-StAtom}. Then the atomic transition probability of Eq. \eqref{eq:Atom-ExProb} will be given by
\begin{eqnarray}\label{eq:Pex-StAtom}
    \mathcal{P}^{ex}_{\varepsilon}(\tilde{\omega}) &=& g^2\bigg|\int_{-\infty}^{\infty}d\tau~ u^{\star}_{\varepsilon\,m\,k} (t,\rho,\varphi,z)\,e^{i\,\tilde{\omega}\,\tau}\bigg|^2~\nonumber\\
    ~&=& \frac{g^2}{4\pi^2 R_{0}^2\,\varepsilon}\, \bigg|\frac{\mathcal{J}_{m}(\kappa_{mn}\,a)}{\mathcal{J}'_{m}(\kappa_{mn}\,R_{0})}\bigg|^2\times \bigg|\int_{-\infty}^{\infty}dT~\, e^{i\,\big(\tilde{\omega}+\frac{\varepsilon}{\gamma}+m\Omega\big)\,T}\bigg|^2~\nonumber\\
    ~&=& \frac{g^2}{R_{0}^2\,\varepsilon}\, \bigg|\frac{\mathcal{J}_{m}(\kappa_{mn}\,a)}{\mathcal{J}'_{m}(\kappa_{mn}\,R_{0})}\bigg|^2\times \Big|\delta\Big[\tilde{\omega}+\Big(m\Omega+\frac{\varepsilon}{\gamma}\Big)\Big]\Big|^2~\nonumber\\
    ~&=& \frac{g^2\,\gamma^2}{R_{0}^2\,\varepsilon}\,\bigg|\frac{\mathcal{J}_{m}(\kappa_{mn}\,a)}{\mathcal{J}'_{m}(\kappa_{mn}\,R_{0})}\bigg|^2\times \Big|\delta\big[\varepsilon+\gamma(m\Omega+\tilde{\omega})\big]\Big|^2~.
\end{eqnarray}
{The above transition probability is dimensionless, like the one obtained from Eq. \eqref{eq:Pex-StMirror}. This becomes evident as the speed of light $c$ is dimensionless, and the square of the coupling constant $(g^2)$ has the dimension of inverse length.}
To obtain the above expression, we have considered the atomic energy gap to be $\tilde{\omega}$. Here also, we would like to mention that the above expression of the atomic transition probability gets much simplified if we identify $(m\Omega+\varepsilon/\gamma)$ as the rotating frame field frequency as observed by a static observer $\varepsilon'$ (see Eq. \eqref{eq:Mode-energy-rel} in this regard). In terms of the field mode frequencies $\varepsilon$ and $\varepsilon'$ of the rotating frame, the above atomic transition probability becomes
\begin{eqnarray}\label{eq:Pex-StAtom-2}
    \mathcal{P}^{ex}_{\varepsilon,\,\varepsilon'}(\tilde{\omega}) &=& \frac{g^2}{R_{0}^2\,\varepsilon}\, \bigg|\frac{\mathcal{J}_{m}(\kappa_{mn}\,a)}{\mathcal{J}'_{m}(\kappa_{mn}\,R_{0})}\bigg|^2\times \big|\delta\big[\tilde{\omega}+\varepsilon'\big]\big|^2~.
\end{eqnarray}
At this stage, we would like to draw attention to the structural similarity between the atomic transition probabilities of Eqs. \eqref{eq:Pex-StMirror-2} and \eqref{eq:Pex-StAtom-2} corresponding to our two different scenarios. It is to be noted that $\chi_{mn}=\kappa_{mn}$, and thus the transition probabilities of Eqs. \eqref{eq:Pex-StMirror-2} and \eqref{eq:Pex-StAtom-2} are equivalent when $\omega \leftrightarrow \tilde{\omega}$, $\mathcal{E} \leftrightarrow \varepsilon$, and $\mathcal{E}' \leftrightarrow \varepsilon'$, i.e., when the field frequencies corresponding to the static and rotating frames interchange.

\subsection{Field mode integrated transition rate}

In this part of the section, we estimate the mode integrated transition probability rate of \eqref{eq:Rex-MdIndp} corresponding to our second scenario. In this regard, we consider the expression of atomic transition probability from the previous expression \eqref{eq:Pex-StAtom}, and with the help of Eq. \eqref{eq:Atom-ExProb-MdIndp}, we obtain the mode integrated transition rate as
\begin{eqnarray}\label{eq:Rex-StAtom1}
    \mathcal{R}^{ex}(\tilde{\omega}) &=& \frac{g^2\gamma^2}{R_{0}^2\,\big\{\lim_{\tilde{T}\to \infty} \, \int_{-\tilde{T}}^{\tilde{T}} d\eta\big\}}\,\sum_{m=-\infty}^{\infty}\sum_{n=1}^{\infty}\int_{-\infty}^{\infty} \frac{dk}{\varepsilon}\, \bigg|\frac{\mathcal{J}_{m}(\kappa_{mn}\,a)}{\mathcal{J}'_{m}(\kappa_{mn}\,R_{0})}\bigg|^2\, \big| \delta[\varepsilon + \gamma(m\Omega+\tilde{\omega})]\big|^2~.
\end{eqnarray}
For a massless scalar field $\mu=0$, we have the relation $(m\Omega+\varepsilon/\gamma)^2=k^2+\kappa^{2}_{mn}$, i.e., one should have $\varepsilon=-\gamma m \Omega+\gamma \sqrt{k^2+\kappa^{2}_{mn}}$ for the above Dirac delta to be non-vanishing. It is to be noted that we considered the positive root of $\varepsilon$ in terms of $\gamma \sqrt{k^2+\kappa^{2}_{mn}}$, as $\varepsilon$ represents the field frequency and cannot be negative. We consider $g(k)=\varepsilon + \gamma(m\Omega + \tilde{\omega})$. Then we may expand the above Dirac delta distribution as
\begin{eqnarray}
    \delta[\varepsilon + \gamma(m\Omega + \tilde{\omega})] &=& \sum_{j} \frac{\delta(k-k_{j})}{|g'(k_{j})|}~,
\end{eqnarray}
where $k_{j}=\pm \sqrt{\tilde{\omega}^2-\kappa^{2}_{mn}}$ and $|g'(k_{j})| = \gamma\,|k_{j}|/\sqrt{k^{2}_{j}+\kappa^{2}_{mn}} = \gamma\, \sqrt{\tilde{\omega}^2-\kappa^2_{mn}}/|\tilde{\omega}|$. Then the previous expression of the transition rate becomes
\begin{eqnarray}\label{eq:Rex-StAtom2}
    \mathcal{R}^{ex}(\tilde{\omega}) &=& \frac{4\,g^2}{R_{0}^2}\,\sum_{m=-\infty}^{\infty}\sum_{n=1}^{\infty}\, \bigg|\frac{\mathcal{J}_{m}(\kappa_{mn}\,a)}{\mathcal{J}'_{m}(\kappa_{mn}\,R_{0})}\bigg|^2\, \frac{(k^2_{j}+\kappa^2_{mn})~\theta(|\tilde{\omega}|-\kappa_{mn})}{\varepsilon_{i}\,|k_{j}|^2}\nonumber\\
    ~&=& \frac{4\,g^2}{R_{0}^2}\,\sum_{m=-\infty}^{\infty}\sum_{n=1}^{\infty}\, \bigg|\frac{\mathcal{J}_{m}(\kappa_{mn}\,a)}{\mathcal{J}'_{m}(\kappa_{mn}\,R_{0})}\bigg|^2\, \frac{\tilde{\omega}^2~\theta(|\tilde{\omega}|-\kappa_{mn})}{\gamma\,(\tilde{\omega}^2-\kappa^2_{mn})\,(|\tilde{\omega}|-m\Omega)}~.
\end{eqnarray}
{The above transition rate has the same dimensions as the rate obtained from Eq. \eqref{eq:Rex-StMirror2}, which is inverse of time. This is consistent with our choice of units, in which the speed of light $c$ is dimensionless, rendering the coupling constant $(g)$ dimensionless as well.}
The above transition probability rate can be estimated numerically for different system parameter values. Moreover, we should bring attention to the fact that the expression for the atomic transition rate \eqref{eq:Rex-StAtom2} is structurally similar to the transition rate of the first scenario \eqref{eq:Rex-StMirror2} up to a $\gamma^2$ factor if one identifies $\tilde{\omega}=m\,\Omega-\omega/\gamma$. We have plotted the transition rate of \eqref{eq:Rex-StAtom2} with respect to atomic radial position $a$ in Fig. \ref{fig:RexVsa-w} and compared it with the first scenario for $\tilde{\omega}=\omega$ and $\tilde{\omega}=m\,\Omega-\omega/\gamma$. Our observations suggest that the transition rates corresponding to the two scenarios are qualitatively similar. 
At the same time, when $\tilde{\omega}=m\,\Omega-\omega/\gamma$ and for small values of $a$, the two scenarios are quantitatively the same as well. Although there is a slight mismatch between them for higher values of $a$. This is also reflected in the second panel of Fig. \ref{fig:RexVsa-w} (notice the difference in the blue dashed and red curves at the second peak). This is expected as the two transition probabilities differ by the Lorentz factor $\gamma$, which is $\gamma\sim 1$ for small values of $a$. From these plots, one can also notice that when $a\to R_{0}$, both the transition rates tend to vanish. It is to be noted that in this limit the quantity $\mathcal{J}_{m}(\kappa_{mn}a)$ vanishes in the expressions of \eqref{eq:Rex-StMirror2}and \eqref{eq:Rex-StAtom2}, giving rise to the observed phenomenon in the plots.

\section{Observations \& Discussion}\label{sec:discussion}

In Mach's principle, an observer in circular motion with respect to a static environment is expected to experience similar classical effects if a static observer is placed inside a rotating environment, and in this work, we aspired to provide a quantum interpretation of it. In this regard, we considered two specific scenarios. In one scenario, an atom is in circular motion inside a static cylindrical boundary. At the same time, in the other scenario, the atom is static inside a cylindrical boundary that is in rotational motion around its axis. We compare the atomic transition probabilities between these two scenarios to understand how the relative motion between the atom and the environment influences the quantum response. 

We obtained the atomic transition probabilities in a situation when the atom interacts with a single field mode, with frequency $\mathcal{E}$ or $\varepsilon$, as well as in a field mode integrated scenario. The key observations from our analysis are as follows.
\begin{itemize}
    \item When the atoms from different scenarios interact with a single field mode, the transition probabilities are obtained in Eqs. \eqref{eq:Pex-StMirror-2} and \eqref{eq:Pex-StAtom-2}. While comparing \eqref{eq:Pex-StMirror-2} and \eqref{eq:Pex-StAtom-2}, one can notice that they are structurally exactly the same, as one can recall that $\chi_{mn} = \kappa_{mn}$. Moreover, if one considers $\mathcal{E} \leftrightarrow \varepsilon$ and $\mathcal{E}' \leftrightarrow \varepsilon'$, i.e., interchanges the static frame mode frequencies as seen by a static and rotating observer by the rotating frame mode frequencies as seen by the rotating and static observer, the atomic transition probabilities become quantitatively the same as well. This observation is reminiscent of Mach's principle \cite{padmanabhan2010gravitation} in terms of atomic transitions.

    \item In regard to the field mode integrated transition probability rates of Eqs. \eqref{eq:Rex-StMirror2} and \eqref{eq:Rex-StAtom2}, we would like to mention that there is a certain structural similarity between the two scenarios when $\tilde{\omega} = m\,\Omega-\omega/\gamma$. This claim is easily visualized in Fig. \ref{fig:RexVsa-w} where we plotted the transition probability rates corresponding to the two scenarios and observed qualitative similarities when  $\tilde{\omega} = m\,\Omega-\omega/\gamma$. 
    {Our observations (Fig. \ref{fig:RexVsa-w}) suggest that, under the condition $\tilde{\omega}=m\Omega-\omega/\gamma$, the two transition rates co-inside with each other for small values of atomic position $a$ and angular velocity $\Omega$. As the linear velocity is $v=a\,\Omega$ the previous observation holds for small values of $v$, i.e., only in the non-relativistic regime, characterized by $v\ll c$. This similarity, however, does not persist beyond the non-relativistic regime.}
\end{itemize}
{Therefore, our observations suggest that in the semi-classical regime, an analogue of the classical Mach's principle exists in terms of the atomic transition probabilities.}
It should also be noted that our conclusions are somewhat different from \cite{DePaola:2000cv}, where a similar objective was pursued with the field modes in the rotating Trocheries-Takeno vacuum state. It would be interesting to see whether any of these predictions can be verified in an actual experimental set-up. 
{In this regard, our present analysis is also relevant, as it is conceptually analogous to the cylindrical cavity configuration, which underlies several intriguing proposals for experimental verification \cite{Stargen:2021vtg, Peng:2024pfy, Arya:2024qke}. Cylindrical cavities are particularly promising in this context because they can enhance atomic responses, thereby offering potential experimental advantages. However, it is worth noting that realizing such a set-up also poses significant experimental challenges, particularly in achieving precise control over the cavity dimensions, potentially down to the nanometer scale.}
{Our findings should also be compared with those for the linear acceleration scenario, in which the quantum equivalence between two configurations -- one in which the atom undergoes uniform acceleration relative to a static mirror and another in which the atom remains static while the mirror accelerates uniformly -- was found to hold only in $(1+1)$ dimensions \cite{Svidzinsky:2018jkp}. In contrast, in $(3+1)$ dimensions, subtle differences in the atomic transition probabilities distinguish the two scenarios \cite{Kumawat:2024kul}.}

\begin{acknowledgments}
SB gratefully acknowledges the financial support provided by the Indian Statistical Institute, Kolkata, through the Visiting Scientist position.

\end{acknowledgments}

\appendix
\section{Transition probability of a static atom inside a cylindrical boundary}\label{Appn:Static-detector}
The transition probability of a static atom can be easily found using the initial expression of atomic transitions leading to Eq. \eqref{eq:Pex-StMirror}. In that expression, we identify $T=\tau$, $R=a$, $\Phi=0$ and an arbitrary $Z$ corresponding to a static atom placed at a distance $a$ from the central axis. Then we get the transition probability as
\begin{eqnarray}\label{Appn:Pex-StAtmMirror}
    \mathcal{P}^{ex}_{\mathcal{E}}(\omega) &=& g^2\bigg|\int_{-\infty}^{\infty}d\tau~ U^{\star}_{\mathcal{E}\,m\,k} (T,R,\Phi,Z)\,e^{i\,\omega\,\tau}\bigg|^2~\nonumber\\
    ~&=& \frac{g^2}{4\pi^2 R_{0}^2\,\mathcal{E}}\,\bigg|\frac{\mathcal{J}_{m}(\chi_{mn}\,a)}{\mathcal{J}'_{m}(\chi_{mn}\,R_{0})}\bigg|^2\times \bigg|\int_{-\infty}^{\infty}d\tau~\,e^{i\,(\omega+\mathcal{E})\,\tau}\bigg|^2~\nonumber\\
    ~&=& \frac{g^2}{R_{0}^2\,\mathcal{E}}\,\bigg|\frac{\mathcal{J}_{m}(\chi_{mn}\,a)}{\mathcal{J}'_{m}(\chi_{mn}\,R_{0})}\bigg|^2\times \big|\delta[\omega+\mathcal{E}]\big|^2~.
\end{eqnarray}
The above expression shows that even for a static atom within a cylindrical boundary, the atomic transition probability exhibits a Dirac delta distribution. In this case, it is evident that the transition probability is non-zero only when $\omega=-\mathcal{E}$, i.e., only the atomic de-excitations are allowed.

\section{Regularization of single mode transition probability for an atom in circular motion}\label{Appn:TP-mode-Regu}
Here, we propose a regularization scheme for the divergences arising in the single-mode atomic transition probabilities. In this regard, one convenient method is to consider finite-time interaction between the atom and the field. Let us consider the interaction to be incorporated through a Gaussian switching function. Then, the interaction Hamiltonian, to begin with, will be modified suitably as 
\begin{eqnarray}\label{eq:Int-Hamiltonian-V2}
    \hat{H}_{I}(\tau) &=& \hbar\,g\,e^{-\tau^2/\varsigma^2}\,\Big(\phi_{\nu}[t(\tau),z(\tau)]\,\hat{a}_{\nu}+h.c.\Big)\,\big(e^{-i\,\omega\,\tau}\hat{\sigma}+h.c.\big)~,
\end{eqnarray}
where $\varsigma$ signifies the effective time scale for the interaction. With this, the single-mode atomic transition probability of Eq. \eqref{eq:Pex-StMirror}, that corresponds to an atom in circular motion and a static mirror, becomes
\begin{eqnarray}\label{Appeq:Pex-StMirror}
    \mathcal{P}^{ex}_{\mathcal{E}}(\omega) &=& g^2\bigg|\int_{-\infty}^{\infty}d\tau~ U^{\star}_{\mathcal{E}\,m\,k} (T,R,\Phi,Z)\,e^{i\,\omega\,\tau}\,e^{-\tau^2/\varsigma^2}\,\bigg|^2~\nonumber\\
    ~&=& \frac{g^2}{4\pi^2 R_{0}^2\,\mathcal{E}}\,\bigg|\frac{\mathcal{J}_{m}(\chi_{mn}\,a)}{\mathcal{J}'_{m}(\chi_{mn}\,R_{0})}\bigg|^2\times \bigg|\int_{-\infty}^{\infty}d\tau~\,e^{i\,(\omega+\mathcal{E}\gamma-m\gamma\Omega)\,\tau}\,e^{-\tau^2/\varsigma^2}\bigg|^2~\nonumber\\
    ~&=& \frac{g^2}{4\pi\,R_{0}^2\,\mathcal{E}}\,\bigg|\frac{\mathcal{J}_{m}(\chi_{mn}\,a)}{\mathcal{J}'_{m}(\chi_{mn}\,R_{0})}\bigg|^2\times \varsigma^2\,e^{-(\omega+\mathcal{E}\gamma-m\gamma\Omega)^2\,\varsigma^2/2}~.
\end{eqnarray}
It is to be noted that the above expression is not divergent. However, in the limit $\varsigma\to\infty$, i.e., in the infinite interaction time limit, one can utilize the limiting expression of the Dirac delta $\delta(z)=\lim_{\alpha\to\infty}=(\alpha/\sqrt{2\pi}) \,\exp\{-\alpha^2\,z^2/2\}$ (see \cite{Barman:2022utm}) and get back the transition probability of expression \eqref{eq:Pex-StMirror}. Here, we would like to mention that in a similar manner, one can also regularize the atomic transition probability of Eq. \eqref{eq:Pex-StAtom} corresponding to a scenario where the atom is static and the mirror is in circular motion.

\bibliographystyle{utphys1.bst}

\bibliography{bibtexfile}

\end{document}